# Magnetochiral Tunneling in Paramagnetic $Co_{1/3}NbS_2$

Seongjoon Lim[1†], Sobhit Singh[2,3†], Fei-Ting Huang[1], Shangke Pan[1,4], Kefeng Wang[1], Jaewook Kim[1], Jinwoong Kim[5], David Vanderbilt[5*], Sang-Wook Cheong[1*]

[1]Rutgers Center for Emergent Materials and Department of Physics and Astronomy, Piscataway, NJ 08854, USA
[2]Department of Mechanical Engineering, University of Rochester, Rochester, NY 14627, USA
[3]Materials Science Program, University of Rochester, Rochester, New York 14627, USA
[4]State Key Laboratory Base of Novel Function Materials and Preparation Science, School of Material Sciences and Chemical Engineering, Ningbo University, Ningbo, Zhejiang 315211, China
[5]Department of Physics and Astronomy, Rutgers University, Piscataway, New Jersey 08854, USA

[†] These authors contributed equally to this work.

*Corresponding Authors: David Vanderbilt, Sang-Wook Cheong
**Email:** dhv@physics.rutgers.edu, sangc@physics.rutgers.edu

**Abstract**
Electric currents have the intriguing ability to induce magnetization in nonmagnetic crystals with sufficiently low crystallographic symmetry. Some associated phenomena include the non-linear anomalous Hall effect in polar crystals and the nonreciprocal directional dichroism in chiral crystals when magnetic fields are applied. In this work, we demonstrate that the same underlying physics is also manifested in the electronic tunneling process between the surface of a nonmagnetic chiral material and a magnetized scanning probe. In the paramagnetic but chiral metallic compound $Co_{1/3}NbS_2$, the magnetization induced by the tunneling current is shown to become detectable by its coupling to the magnetization of the tip itself. This results in a contrast across different chiral domains, achieving atomic-scale spatial resolution of structural chirality. To support the proposed mechanism, we used first-principles theory to compute the chirality-dependent current-induced magnetization and Berry curvature in the bulk of the material. Our demonstration of this magnetochiral tunneling effect opens up a new avenue for investigating atomic-scale variations in the local crystallographic symmetry and electronic structure across the structural domain boundaries of low-symmetry nonmagnetic crystals.

**Significance Statement**
The prevailing assumption in scanning tunneling microscopy (STM) has been that the magnetic signal originates solely from an interplay between the magnetizations of the tip and sample. Here, we report a novel use of magnetic STM to detect a broken crystallographic symmetry in the absence of magnetic order in the sample. Specifically, we show that magnetic STM applied to nonmagnetic $Co_{1/3}NbS_2$ can distinguish the two chirally reversed surface domains via the magnetization induced in the sample by the tunneling current, an effect that is facilitated by the presence of Berry curvature in the electronic band structure. Our findings open new opportunities for using magnetic STM to provide high-resolution images of nonmagnetic domain structures associated with purely crystallographic broken symmetries.

**Introduction**

Since its invention, scanning tunneling microscopy (STM) has had a profound impact on the fields of nanotechnology and materials science due to its unique ability to achieve atomic-scale spatial and meV energy resolution, and to manipulate atoms (1). Its magnetic variant, spin-polarized scanning tunneling microscopy (SP-STM), which makes use of magnetized probes, has been particularly useful in the study of atomic-scale magnetic order (2). Notably, SP-STM has been extensively used to explore such novel phenomena as Majorana zero modes (3) and single-molecule magnets (4). However, to date the application of SP-STM has primarily been limited to the study of spontaneous long-range magnetic order in the sample, since it relies on distinct tunneling probabilities for parallel and antiparallel magnetizations between probe and sample (2).

The recent discovery of current-induced bulk magnetization in nonmagnetic chiral structures (5–10) has prompted us to consider the feasibility of observing current-induced magnetization through the application of SP-STM. While tunneling currents are small, they become highly concentrated in the sample just below the tip, so that the magnetic detection of nonmagnetic domains with opposite chiralities at a crystal surface becomes plausible. Such an application of SP-STM promises to provide enormously improved resolution for imaging of chiral domain structures compared to diffraction-limited optical techniques.

Moreover, the phenomenon of current-induced bulk magnetization, known as the kinetic magnetoelectric effect, has been investigated theoretically (11–15) and shown to be closely related to some unconventional quantum responses such as nonlinear anomalous Hall (16–20), second harmonic transport (21), kinetic Faraday (12, 22), gyrotropic magnetic (11), and circular photogalvanic (12) effects, as well as the Edelstein effect in 2D (23). These have been shown to be governed by quantum geometric effects associated with the presence of Berry curvature in the electronic band structure of acentric crystals even without broken time-reversal symmetry. In particular, there has been considerable recent interest in the so-called "Berry curvature dipole" (BCD), which describes an imbalance of Berry curvature in the Brillouin zone and governs the nonlinear anomalous Hall effect (17, 19, 24–26). Theory indicates an intimate connection between the current-induced spin and orbital magnetization and nonlinear Hall effects, which also have very similar broken symmetry requirements and typically appear together in chiral crystals (24, 25, 27). In fact, these properties can be used as a probe of broken spatial symmetries arising from chiral, polar, ferro-rotational, or other structural orders, based either on conventional symmetry analysis or in the framework of symmetry operational similarity (28). They offer an intriguing opportunity to explore the emergence of quantum phenomena arising from Berry curvature, especially those associated with topological features of the band structure (13), at the atomic scale if they can be coupled to a local magnetic probe such as a scanning tip.

In this work, we demonstrate the proposed SP-STM detection of current-induced magnetization to reveal crystallographic chirality in nonmagnetic substrates. Specifically, we show that $Co_{1/3}NbS_2$, a structurally chiral nonmagnetic material, can host a type of kinetic magnetoelectric effect in which an electric current tunneling out of the surface and into the tip carries a detectable magnetization, providing a contrast mechanism for mapping chiral domain structures. We will use the term "magnetochiral tunneling effect" (MTE) for this novel magnetoelectric phenomenon. Our symmetry considerations and first-principles calculations of the uniform bulk current-induced spin and orbital magnetizations and BCD provide support for the conclusion that the observed magnetic signal is, indeed, caused by these phenomena operating in the region of high current density concentrated under the scanning tip. This finding provides a novel mechanism for the direct determination of crystallographic chirality at the atomic scale, highlights the role of electronic Berry curvature in this novel phenomenon, and opens the possibility of studying other types of broken crystallographic symmetries using a similar approach.

**Results**

In order to detect the MTE, we choose a class of layered two-dimensional chiral metals, $M_{1/3}(Nb,Ta)S_2$ (M=3$d$ transition metals), which have attracted significant interest due to the structural chiral domains outlined by a network of vortex and anti-vortex pairs (29–31). The sub-micron-scale chiral domains and boundaries that comprise the heterochiral structure in $M_{1/3}(Nb,Ta)S_2$ provide a unique platform for observing the influence of quantum geometric effects on the opposite structural chiralities. Additionally, $Cr_{1/3}NbS_2$, one of the isostructural materials with helical magnetism, has been reported to show chirality-induced magnetic transport (9). Therefore, even in the absence of magnetic order, it might be possible to find a similar chirality-induced magnetoelectric interaction in an isostructural member. Focusing on $Co_{1/3}NbS_2$, we use SP-STM to detect the change in the magnetic signal produced by the magnetoelectric interaction across a chiral domain boundary at the atomic scale. We further elucidate the origin of the observed magnetic signal across the chiral domain boundary based on first-principles density-functional theory (DFT) calculations that demonstrate the appearance of non-vanishing magnetization and net Berry curvature under a quasi-equilibrium flow of electric current.

**Symmetry Considerations:** First, we examine the symmetry operational (28, 32) properties of various components related to MTE or the current-induced magnetization (5, 9, 33, 34) in the bulk chiral structure. Fig. 1A illustrates that the presence of a mirror plane parallel to the electric current switches the chirality (green helix) as well as the sign of the induced magnetization (blue arrow), while the electric current (orange arrow) remains unchanged. On the other hand, a two-fold rotation ($C_2$) around an axis perpendicular to the current reverses only the magnetization and the current, not the chirality. This relationship, reminiscent of a magnetic field induced by a current flowing around a coil, can also be examined in reciprocal space in relation to Berry curvature (Fig. 1B). In the presence of an electric current (orange arrows), the Berry curvature (small light-blue arrows) along the direction of the flow determines the induced magnetization (blue arrows). The Berry curvature and magnetization, being pseudo-vectorial quantities, change sign under the mirror operation. Instead, application of a $C_2$ rotation around an axis perpendicular to the current flips the electric current direction and the induced magnetization, while the texture of the Berry curvature remains unchanged. Therefore, the symmetry analysis predicts that the current-induced MTE should be reversible by two factors: structural chirality and electric current direction (28, 32). Further analysis based on symmetry operational similarity suggests that other types of spatial symmetries, such as polar structures, can also exhibit similar coupling between electric current and the broken lattice symmetry. We provide more details in the supplementary information (SI note 1).

**STM/SP-STM Results:** The structure of $Co_{1/3}NbS_2$ depicted in Fig. 1C is formed through the intercalation of Co ions into the van der Waals gap of the centrosymmetric parent material 2H-$NbS_2$. The structure belongs to one of the 65 Sohncke groups (SG#182, $P6_322$) that can exhibit opposite chiral structures within the same space group (35). The Co ions within a van der Waals gap occupy one of the three Nb-atop sites, which are labeled A, B, and C. The intercalation sites in adjacent gaps along the $c$ direction alternate between two sites. Since only two of the three Nb-atom sites are occupied, there are six permutation pairs (*i.e.* AB, BC, CA, BA, CB, and AC). The six pairs form two kinds of structural chiralities, which we designate as + and – chirality (30). Within each chirality, there are three structures that are mutually convertible by $C_3$ rotation (AB, BC, and CA for + chirality; BA, CB, and AC for – chirality). The formation of a topological vortex showing alternating chiral domains is described in Fig. 1D, and the real-space observation using dark-field transmission electron microscopy is shown in Fig. 1E. The present result is the first report of the formation of a topological vortex in $Co_{1/3}NbS_2$, but it is consistent with previous observations of topological vortex formation in the isostructural $M_{1/3}TaS_2$ compounds (29, 30, 36). We note that there are two types of boundaries depending on the Co layer in which the lattice shifts, which are depicted in Fig. 1D with solid and dashed lines. The boundaries with solid lines, which involve a change in the first character (e.g. AB to CB), are accompanied by a change in the intercalation site in the upper Co lattice (i.e. the exposed Co lattice on a surface). The boundaries with dashed lines (e.g. CB to CA) involve a change in the lower Co lattice (i.e. the Co underneath the $NbS_2$ layer).

We cleave samples for STM and SP-STM at a cleaving stage with liquid nitrogen cooling (31) and all measurements are carried out at 78 K to avoid the antiferromagnetic ordering, which occurs below 29 K (37–39) (Fig. S3 for susceptibility measurement). We adopt liquid nitrogen cooling to achieve optimal thermal stability at a temperature above the antiferromagnetic order. While it constrains the use of helium-cooled superconducting coil, the extended holding time of liquid nitrogen facilitates the consistent application of the lateral coarse-moving mechanism of the STM setup until a topological vortex is pinpointed. On cleaved surfaces, we could identify two types of surfaces that expose Co and $NbS_2$ on the top surface (Fig. S5), and only Co-type surfaces are adopted in this study. The topographies in Fig. 2A and B highlight the observation of a topological vortex in STM and SP-STM, respectively. The STM result clearly reveals the three domain boundaries connected to the vortex despite the presence of various atomic defects such as Co ions in a metastable charge state (bright spots) and Co vacancies (dark spots), which we have identified in our previous study (31). On the other hand, SP-STM on the same surface region reveals an additional two-level domain contrast around the vortex, with the contrast being more pronounced in a larger-scale image due to averaging out of randomly distributed atomic defects (Fig. 2B inset).

To understand the appearance of only three domain boundaries instead of six in the STM image, we set up two atomic models of the chiral domain boundaries as shown in Fig. 1D based on our previous study (30). Fig. 2F and 2G show the models with a shift in the Co lattice above and underneath the $NbS_2$ layer, respectively. The difference is that Fig. 2F has a shift of the surface Co lattice (circles with solid lines) by 1/3 of the unit cell size, while Fig. 2G does not. Analysis of the STM image (Fig. 2C) shows that the three domain boundaries in the STM image belong to the type shown in Fig. 2F. The other three boundaries underneath could not be located within the bias range of a few hundreds of mV, and use of higher bias voltage is limited by the presence of metastable charge states of Co ions(31).

In stark contrast to the STM results, the two-level domain contrast appears in the SP-STM image, facilitating the identification of all six chiral domains and boundaries. Among the six chiral domain boundaries, only three of them showed a shift of Co lattice (Fig. 2D) similar to the STM result. The other three boundaries, on the other hand, do not show a shift (Fig. 2E), which is consistent with the second type domain boundary (Fig. 2G). Instead, there exists a stepwise change in topography as shown in Fig. 2H. Because the contrast is only observed when using a magnetized probe, we assign the signal to be magnetic suggesting that it is related to a magnetoelectric interaction with the structural chirality. We exclude the possibility that this two-level contrast originates from a long-range order of Co spins, as the measurement temperature (78K) is well above the antiferromagnetic transition temperature (29K) (37–39). Furthermore, the two-level domain contrast that is locked to the structural chirality is not consistent with the stripe pattern expected from the antiferromagnetic order (40). Thus, we explore the possibility that the magnetic two-level contrast is related to the MTE that can occur in chiral structures (5, 6, 10, 12, 15, 33, 41), using the formalism of gyrotropic magnetoelectric and BCD effects in the electronic structure (19, 24, 26, 42) as we explain next.

**Theoretical Analysis:** We use density-functional theory (DFT+U) calculations with spin-orbit coupling (SOC) to analyze the electronic band structure of the two inverted bulk chiral structures of $Co_{1/3}NbS_2$. We carry out DFT+U+SOC calculations (see Methods) for ferromagnetic, antiferromagnetic, and nonmagnetic configurations (SI note 9), obtaining a metallic ground state with a complex Fermi surface dominated by Nb $d$, Co $d$, and S $p$ orbitals in all cases. Since the material is paramagnetic at the experimental conditions, we focus henceforth on the nonmagnetic case, for which the band structure is shown in Fig. 3A. For consistency, we continue to use DFT+U (43), although the Hubbard $U$ has very little effect on the nonmagnetic calculation (SI note 10).

To better understand how a current flowing along the $z$ symmetry axis of $Co_{1/3}NbS_2$ activates the appearance of a corresponding $z$ magnetization, we examine the $z$ components of the spin ($S_z$) and Berry curvature ($\Omega_z$) in reciprocal space. Fig. 3B shows the expectation value of $S_z$ for bands along the Γ-A ($k_z$) high-symmetry direction of the Brillouin zone (BZ) for both chiral variants. Fig. 3C presents $S_z$ projected onto the Fermi surface, while 3D shows $S_z$ projected onto the Fermi loops on a $k_x$-$k_z$ plane at $k_y = 0$. We confirm that both $S_z$ and $\Omega_z$ change sign upon chirality reversal. We also note that they reverse sign under $\mathbf{k} \to -\mathbf{k}$ in reciprocal space, i.e., $\Omega_z(\mathbf{k}) = -\Omega_z(\mathbf{k})$ and $S_z(\mathbf{k}) = -S_z(\mathbf{k})$, as expected from time-reversal symmetry. The orbital angular momentum $L_z(\mathbf{k})$ (not shown) follows the same symmetry rules. Each of these three quantities integrates to zero over the occupied Fermi sea at equilibrium, but becomes non-zero because of the imbalanced occupation that results from a flow of electric current along $z$ as schematically shown in Figs. 4A-B. (7, 19). This is shown for the Berry curvature in Fig. 3C, where the green arrows indicate the direction of the BCD, but the same applies to the generation of magnetization. Since we do not know the details of the tunneling process from sample to tip, we remain open to the possibility that all of these induced quantities – spin, orbital moment, and Berry curvature – could contribute in different ways.

To be specific, we compute the two kinetic magnetoelectric tensors $K^{\text{spin}}$ and $K^{\text{orb}}$ and the BCD tensor $D$ defined by

$$K^{\text{spin}}_{ij} = \sum_n \frac{1}{(2\pi)^3} \int d^3k\, f_{n\mathbf{k}} \frac{\partial}{\partial k_j} m^{\text{spin}}_{n\mathbf{k},i}, \qquad \ldots (1)$$

$$K^{\text{orb}}_{ij} = \sum_n \frac{1}{(2\pi)^3} \int d^3k\, f_{n\mathbf{k}} \frac{\partial}{\partial k_j} m^{\text{orb}}_{n\mathbf{k},i}, \qquad \ldots (2)$$

$$D_{ij} = \sum_n \frac{1}{(2\pi)^3} \int d^3k\, f_{n\mathbf{k}} \frac{\partial}{\partial k_j} \Omega_{n\mathbf{k},i}. \qquad \ldots (3)$$

These equations describe the spin magnetization, orbital magnetization, and integrated Berry curvature induced by a uniform displacement of the Fermi surface along direction $j$, where $m^{\text{spin}}_{n\mathbf{k},i}$ and $m^{\text{orb}}_{n\mathbf{k},i}$ are the $i^{th}$ components of the spin and orbital moments of the Bloch state of band $n$ at wavevector $\mathbf{k}$ with occupation $f_{n\mathbf{k}}$, and $\Omega_{n\mathbf{k},i}$ is the corresponding Berry curvature. The responses to a static uniform electric field $E_j$ in the limit of a constant relaxation time $\tau$ are given by $M_i = \chi_{ij} E_j$ with $\chi^{\text{spin}}_{ij} = (-e\tau/\hbar) K^{\text{spin}}_{ij}$ and $\chi^{\text{orb}}_{ij} = (-e\tau/\hbar) K^{\text{orb}}_{ij}$, while the nonlinear Hall effect is described by an induced current $j_i = \chi^{\text{NLH}}_{ijk} E_j E_k$ with $\chi_{ijk}^{\text{NLH}} = (-e^3\tau/2)\epsilon_{ilk}D_{jl}$, in implied sum notation ($e > 0$ is the charge quantum, and $\hbar$ is the reduced Planck's constant).

Under the $D_6$ point group, the kinetic magnetic tensors have components $K_{xx} = K_{yy}$ and $K_{zz}$ as the only allowed non-zero elements, while in addition $D_{xx} = D_{yy} = -D_{zz}/2$ for the special case of the BCD (12). It is straightforward to show that these tensors transform in the usual way under proper crystal symmetry operations, but acquire an extra minus sign under improper ones, thus explaining the sign reversal for the case of opposite chiral domains.

Since experimentally our surface is normal to $z$ and the tip bias generates an electric field that is dominantly along $z$ in the region just below the STM tip, we now focus on computing the $zz$ components of these tensors. In practice we use the Fermi-surface formulation obtained from an integration by parts on Eqs. (1-3), i.e.,

$$K_{zz}^{\text{spin}}(E_F) = \sum_n \frac{1}{(2\pi)^3} \int d^3k \, \frac{\partial E_{n,k}}{\partial k_z} \, m_{nk,z}^{\text{spin}} \, \delta(E_{nk} - E_F), \quad \ldots\ldots (4)$$

and similarly for $K_{zz}^{\text{orb}}$ and $D_{zz}$ (11, 12). We also consider variations of $E_F$ in these equations as describing, in a rigid-band approximation, variable bias voltages that may be applied in the tunneling experiments.

In Figure 4C (upper), we plot the variation of $K_{zz}^{\text{orb}}$ and $K_{zz}^{\text{spin}}$ calculated as a function of the chemical potential relative to the Fermi energy. Our calculations reveal that the $K_{zz}^{\text{orb}}$ contribution in Co$_{1/3}$NbS$_2$ is about two orders of magnitude larger than the $K_{zz}^{\text{spin}}$ contribution. This is consistent with the weakness of SOC in this compound, and the fact that $K_{zz}^{\text{spin}}$ would vanish in the absence of SOC while $K_{zz}^{\text{orb}}$ (and the BCD) would not. This is the reverse of the usual situation in static magnetic compounds, where spin effects typically dominate; the difference here is that the tunneling current, which drives these effects, is itself orbital in nature. We observe that both the $K_{zz}^{\text{spin}}$ and $K_{zz}^{\text{orb}}$ as well as $D_{zz}$ (see SI) reverse their sign upon the chirality reversal. Consequently, the current-induced magnetization resulting from these quantities will also reverse its direction, in agreement with our experimental findings.

Going one step further, we estimate the magnitude of the current-induced magnetization using the calculated spin and orbital $K_{zz}$ data. We find it convenient to introduce the tensors $\tilde{\chi}_{ij}^{\text{spin}} = dM_i^{\text{spin}}/dJ_j$ and $\tilde{\chi}_{ij}^{\text{orb}} = dM_i^{\text{orb}}/dJ_j$, which have units of length and characterize the magnetizations in directions $i$ induced per unit current along $j$. Making use of the Ohmic conductivity $\sigma_{ij} = dJ_i/dE_j$ and taking note of the presence of the symmetry axis, it follows that $\tilde{\chi}_{zz}^{\text{spin}} = \chi_{zz}^{spin}/\sigma_{zz} = (-e\tau/\hbar\sigma_{zz})K_{zz}^{spin}$, and similarly for $\tilde{\chi}_{ij}^{\text{orb}}$. We illustrate the variation of $\tilde{\chi}_{zz}$ as a function of the chemical potential near the Fermi energy in Fig. 4C (lower). At the Fermi energy, we find $\tilde{\chi}_{zz}^{\text{orb}}$ = -0.7 Å whereas $\tilde{\chi}_{zz}^{\text{spin}} = 0.002$ Å. The relatively small value of the latter (and of $K_{zz}^{\text{spin}}$) is due to the moderate SOC of the constituent elements in $Co_{1/3}NbS_2$.

In addition, we evaluate the total bias-dependent magnetization density $M_z$ by taking the experimentally measured $J_z$ at each bias voltage and multiplying by the $\tilde{\chi}_{zz}^{\text{total}}$ computed at that bias. The results are plotted in Fig. 4D after converting units from A/cm to $\mu_B$/cell. The maximum estimated $|M_z|$ is about $3\times10^{-3}$ $\mu_B$/cell at a bias voltage of −50 mV. We note that the same order of magnitude of current-induced $M_z$ response has been experimentally observed in the isostructural chiral compound $Cr_{1/3}NbS_2$ by SQUID magnetometry (8). The derivation and calculation of $\tilde{\chi}_{zz}$ along with the theoretical assessment of current-induced magnetization $M_z$ constitute pivotal outcomes of this study.

**Spin-Sensitive Tunneling Spectroscopy:** To verify the behavior of the induced magnetization, we compared the experimental tunneling spectra obtained using a magnetic probe from two chiral domains. The red and blue curves in Fig. 4E show the tunneling spectra obtained on top of the Co atoms in opposite chiral domains shown with red and blue stars in Fig. 4F, respectively. Since the absolute value of the tunneling current depends on the tip height, we carefully compared spectra taken with normalization biases as low as 1 mV to reduce the uncertainty in the height. Fig. 4E shows the averaged results between +1 and −1 mV normalization bias. The comparison confirmed that the crossing point is located at zero bias within the experimental resolution, with an antisymmetric behavior as in the calculated $M_z$ in Fig. 4D. The inset of Fig. 4E depicts the difference after removing the chirality-averaged signal with up to 10% of deviation from the average. The detected asymmetry in the tunneling spectra aligns with estimation of the spin polarization in the tunneling current as outlined in the supplementary information (SI note 4). However, there are still matters to be explored such as a detailed characterization of tip magnetization (44) and or the extent of coupling between orbital angular momentum spin-polarization of the tip (45). We emphasize that, in conventional SP-STM measurements, the crossing points between the opposite magnetic signals usually occur at an arbitrary bias voltage where the dominance of majority and minority spin states changes (2). We consider the zero-bias crossing in our observation to be a critical difference from conventional SP-STM experiments.

Figs. 4A and B schematically explain the mechanism of MTE in relation to non-zero BCD. Although the BCD has opposite signs in the opposite chiral structures, that does not induce any net real-space magnetization in the absence of an electric current (Fig. 4A), because the positive and negative contributions in k-space cancel each other. However, when a flow of electric current is introduced by tunneling, the shift of the Fermi surface breaks the balance of Berry curvature and leads to a dominance from one side. The + chiral structure has a stronger contribution of positive Berry curvature (Fig. 4B, upper), while the – chiral structure has a more negative contribution (Fig. 4B, lower). The magnetic moments $m_{nk,z}$, which follow a similar pattern, result in opposite magnetizations $M_z$ in real space; these can be detected by the magnetized probe through enhanced (parallel magnetizations) or suppressed (antiparallel magnetizations) tunneling probability.

Lastly, we trace the change of the antisymmetric behavior continuously across a domain boundary in Fig. 4F. The magnetic tunneling spectra show the atomic corrugation of twelve Co atoms in the path. While the spectral intensity of the left domain is enhanced at negative bias, it is enhanced in the right domain at positive bias. Again, this antisymmetric behavior indicates that the effect vanishes at zero bias, similar to the other nonequilibrium Berry physics phenomena such as the nonlinear Hall effect (16–20, 46) and the kinetic magnetoelectric effect (12).

**Conclusion:** We have used SP-STM to demonstrate the detection of crystallographic chirality without long-range magnetic order of Co spins in $Co_{1/3}NbS_2$. By comparing observations made with non-magnetic and magnetic probes, we have found a change in the magnetic signal across a chiral domain boundary that is correlated with the structural chirality. With the support of comprehensive theoretical calculations, we attribute these observations, which we refer to as a magnetochiral tunneling effect, to the generation of net magnetization and Berry curvature in the region of concentrated current under the tip. The sign reversal of the experimental signal with crystallographic chirality and with the direction of the tunneling current is consistent with a symmetry analysis. From a fundamental point of view, our work highlights novel ways in which the distributions of Berry curvature and magnetic moments in reciprocal space can manifest themselves as atomic-scale phenomena in real space, while providing, on the practical side, a new tool that delivers unprecedented resolution in the imaging of chiral domain structures.

**Materials and Methods**

**Crystal growth**

Single crystals of $Co_{1/3}NbS_2$ were grown by a chemical vapor transport reaction method in the presence of iodine as a transport agent. About 0.2 g mixture of cobalt powder (Alfa Aesar, 99.99%), niobium powder (Alfa Aesar, 99.9%), and sulfur piece (Alfa Aesar, 99.999%) with a molar ratio Co:Nb:S=0.7:1:2 were sealed in a quartz tube with 100 mg of iodine (Alfa Aesar, 99.99%) under vacuum. Then, the quartz tube was placed under an optimum temperature gradient in a two-zone tube furnace. The hot and cold zones were kept at 950°C and 800°C, respectively. After about 4 weeks, the tube furnace was turned off and the quartz tube was cooled to room temperature naturally. Crystals in hexagonal shape were collected, and X-ray diffraction data on ground powder specimens were taken at room temperature with Cu $K_\alpha$ ($\lambda = 0.15418 nm$) radiation in a Malvern Panalytical X'Pert 3 powder diffractometer.

**Transmission Electron Microscopy**

Specimens for TEM experiments were prepared by Ar-ion milling and studied using a JEOL-2010F TEM. To unveil the chiral domains, DF-TEM images were taken by selecting $g\pm = \pm 22\bar{2}$ along the [101] zone axis. All images are raw data.

**Scanning Tunneling Microscopy**

STM measurements were performed using a Unisoku ultra-high vacuum SPM system (USM-1500) equipped with a cleaving stage capable of cryogenic cooling. Cu(111) and Cr(001) samples cleaned by repeated cycles of sputtering and annealing have been used as reference samples as well as a tip treatment base. Pt/Ir alloy tips were prepared by electron bombardment heating in ultra-high vacuum and treated separately on Cu(111) and Cr(001) surface to be used for normal STM and SP-STM. We used tips that are characterized on Cu(111) surface prior to the measurement for STM. For SP-STM measurements, we utilized Cr(001) surface to coat tips and the magnetization of the tips are confirmed by observing the characteristic layer-by-layer antiferromagnetic contrast of Cr(001) surface. The treatment was done by the tip shaping mechanism implemented in Nanonis controller typically with indentation into the surface with a negative bias to the tip and pulling out with a few nm/s of speed. Several repetitions of the procedure result in Cr transfer to the tip. $Co_{1/3}NbS_2$ sample is glued to a sample plate by silver epoxy (Epotek H20E) and a metal post is attached to the top with the same epoxy. Then the sample is cleaved at 80 K in the cleaving stage after precooling of 1 hour. Coarse moving mechanism of STM system is heavily used to locate a vortex by following a chiral domain boundary. All the STM and SP-STM measurements are acquired at 78 K. The tunneling spectroscopy measurement is obtained by modulation of bias and demodulation of tunneling current using lock-in technique.

**DFT Calculations**

First-principles density-functional theory (DFT+U) calculations were performed using the Vienna ab initio simulation package (VASP) within the projected-augmented wave (PAW) framework (47, 48). We considered nine, eleven, and six valence electrons in the PAW pseudopotential of Co, Nb, and S, respectively. We used the PBEsol exchange-correlation functional for solids (49) along with the rotationally invariant DFT+U method introduced by Liechtenstein et al. (50) to treat the on-site Coulomb correlation effects of the Co 3d electrons at the mean field level. $U = 5$ eV and $J = 1$ eV parameters were considered for Co 3d states in our DFT+U calculations. These values of the ($U$, $J$) parameters correctly predict the ground state magnetic configuration of $Co_{1/3}NbS_2$ (43). We did not observe any substantial change in the electronic properties on tuning $U$ in the 3-6 eV range. A non-magnetic configuration was considered in the reported results; however, ferromagnetic and antiferromagnetic orderings yielded similar metallic features with a complex Fermi surface in $Co_{1/3}NbS_2$. A Γ-centered $k$ mesh of size $12 \times 12 \times 6$ was used to sample the reciprocal space, and 650 eV was used as the kinetic energy cutoff of the plane wave basis set. The structures were optimized until the residual Hellmann-Feynman forces on each atom were less than $10^{-4}$ eV/Å, and $10^{-8}$ eV was defined as the convergence criterion for the electronic self-consistent calculations. Both chiral $Co_{1/3}NbS_2$ structures belong to space group $P6_322$ (SG#182) with the DFT+U optimized lattice parameters $a = b = 5.692$ Å and $c = 11.531$ Å, which are in good agreement with the previously reported experimental data (43). We refer to the AB structure with Co inner coordinates (1/3, 2/3, 1/4) and (2/3, 1/3, 3/4) as the Chiral +, and the inverted BA structure with Co inner coordinates (2/3, 1/3, 1/4) and (1/3, 2/3, 3/4) as the Chiral – structure. SOC was considered in all DFT+U calculations. The PyProcar package (51) was used for the post-processing of electronic structure data. The WannierBerri package (52) was employed to compute the Berry curvature, Berry curvature dipole, ohmic conductivity, and MTE tensors using the real space tight-binding Hamiltonian generated using Wannier90 (53). The $K_{zz}^{orb}$, $K_{zz}^{spin}$, and Berry curvature dipole tensors were computed using a dense k-grid of size 203 x 203 x 87 with a broadening width of 100 K. Tests indicate that the results are converged to within ~2% with respect to the k-point sampling. We further utilized a recursive adaptive k-mesh refinement scheme to achieve faster convergence of Berry-curvature-related properties. A relaxation time of $\tau = 10^{-12}$ s was adopted.

**STM Simulations**

STM simulations in the SI were performed using the GPAW package (54) within the PAW formalism implemented in the Atomic Simulation Environment (55) (ASE). We used the PBE exchange-correlation functional (56) with SCF convergence criteria of $4.0 \times 10^{-8} \, eV^2/el.$ (integrated residual of Kohn-Sham equation), $5.0 \times 10^{-4}$ eV (energy), and $10^{-4}$ electrons/valence electron (electron density). A Γ-centered k-point mesh of size 2x2x2 was used to sample the $k$-space, and a 0.2 Å real space-grid was used for wavefunction expansion in the finite-difference scheme. The surface of $Co_{1/3}NbS_2$ was modeled with four slabs of $2H$-$NbS_2$ with embedded intercalants in between. The top-most Co lattice was added/removed according to the type of surface being simulated, with 10 Å of vacuum added on top. The STM simulations were performed using the local density of states obtained from density functional theory calculations following the Tersoff-Hamann approach (57) as implemented in the GPAW package.

**Acknowledgments**: This work was supported by the W. M. Keck foundation grant to the Keck Center for Quantum Magnetism at Rutgers University. DV acknowledges support from NSF Grant DMR-1954856. SS was supported by the U.S. Department of Energy, Office of Science, Office of Fusion Energy Sciences, Quantum Information Science program under Award Number DE-SC-0020340. SS also acknowledges support from the University Research Awards at the University of Rochester.

**Author Contributions:** S.C. initiated the study; S.L. performed STM experiment and simulation; S.S. performed first-principles calculations; F.H. performed the TEM; S.P. and K.W. grew the crystals; Jaewook Kim performed susceptibility measurement; S.S., Jinwoong Kim, S.L., and D.V. discussed the theoretical results; S.L., S.S., D.V., and S.C. analyzed the data and wrote the paper.

**Competing Interest Statement:** The authors declare that they have no conflict of interest.

# Figures

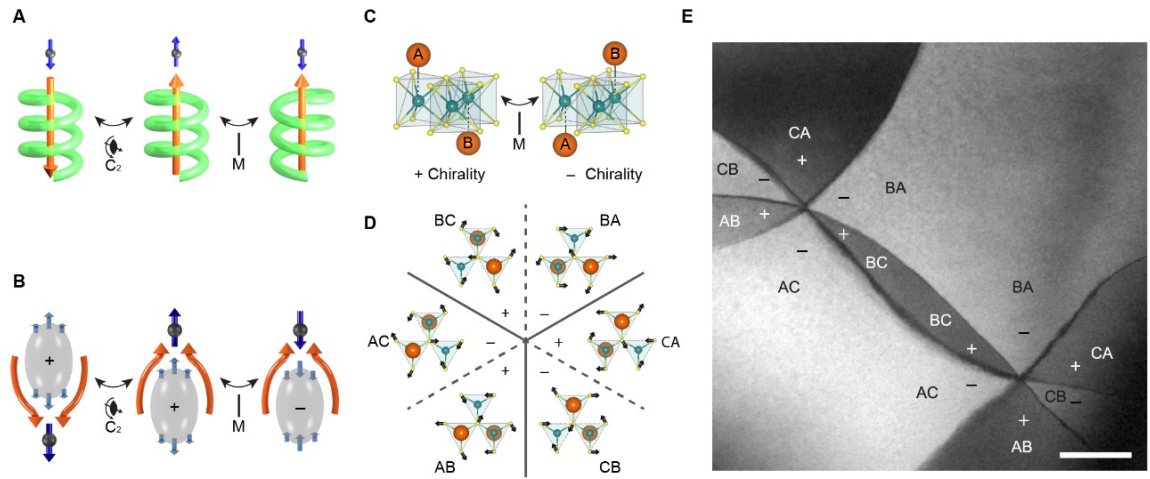

**Figure 1. Mirror Symmetry Operation on (Pseudo) Vectors and Chiral Structures. (A,B)** Changes of (pseudo) vectorial quantities under $C_2$ and mirror operations in real (A) and reciprocal (B) space. Chiral structure (green helix), electric current (orange arrow), magnetization (blue arrow), and Berry curvature (small light-blue arrow) are transformed by mirror operation (right) and $C_2$ rotation (left) from the reference (center). The structural chirality is indicated by + and – signs in (B). **(C)** Two chiralities (+ and –) in the chiral layered compound $Co_{1/3}NbS_2$ are connected by a mirror symmetry operation. The alternation of the intercalant sites of Co atoms (orange) along the *c* axis gives AB and BA type structures with opposite chirality. **(D)** Six configurations of antiphase domains with alternating structural chirality around a vortex. Solid (dashed) line indicates a change of Co lattice above (below) the $NbS_2$ layer. **(E)** Topological vortex and antivortex pair with six domains observed with transmission electron microscopy. The dark and bright contrast reveals the structural chirality as indicated by + and – signs, respectively (scale bar = 0.5 μm).

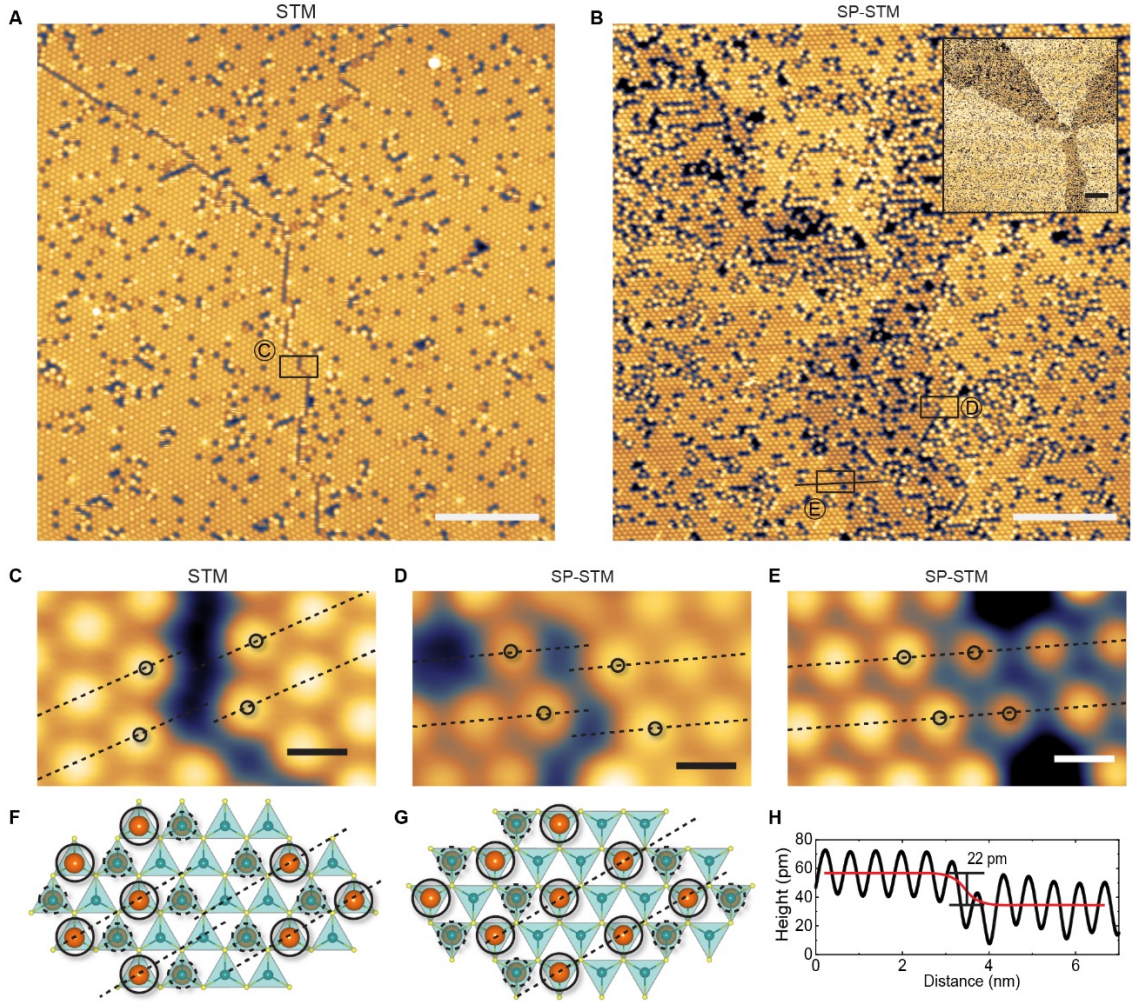

**Figure 2. STM and SP-STM Observation of Chiral Domains around a Vortex (A)** STM topography shows only three domain boundaries around the topological vortex at the center (5mV, 5pA, scale bar = 10 nm). **(B)** SP-STM observation reveals all of six domains by alternating domain contrast. The two-level contrast is more evident in the large-area image in the inset (-50mV, 10pA, scale bar = 10 nm, inset = 20 nm). **(C,D)** Domain boundaries with a shift of the Co lattice on top of the NbS$_2$ layer can be observed in STM (C) and SP-STM (D). The dashed lines show a shift of 1/3 of the unit cell. **(E)** Domain boundaries without a shift of the Co lattice can only be observed in SP-STM. The dashed lines show no shift of the Co lattice (scale bars = 0.7 nm). **(F)** Atomic model of a domain boundary with Co lattice shift above the NbS$_2$ layer that corresponds to (C) and (D). **(G)** Atomic model of a domain boundary with the Co lattice shift underneath the NbS$_2$ layer that corresponds to (E). **(H)** Domain contrast across the domain boundary in SP-STM. The height profile is obtained along the black line in (B) and reveals 22 pm of step-like change across the boundary. The red line is a guide to the eye.

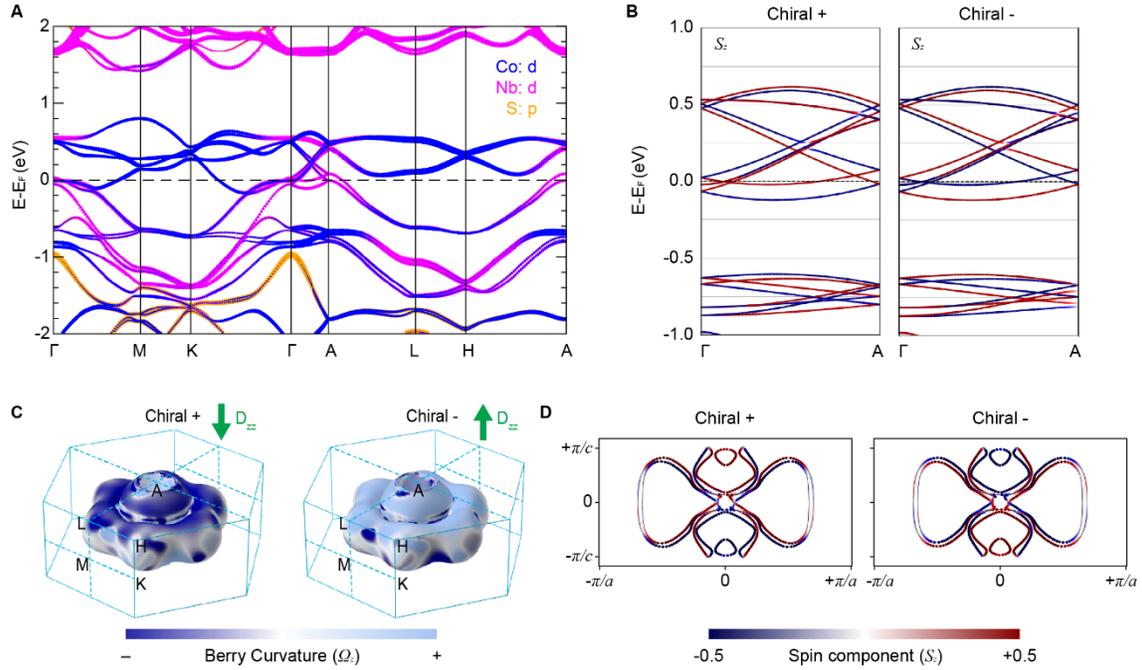

**Figure 3. Electronic Band Structure, Spin Texture, and Berry Curvature. (A)** Atomic-orbital resolved electronic band structure calculated with inclusion of SOC for non-magnetic $Co_{1/3}NbS_2$. **(B)** $S_z$ component of spin magnetization projected on the electronic band structure along the high-symmetry $k_z$ direction of the BZ for + and – chiral structures. Dashed horizontal line depicts the Fermi level. Red and blue colors denote the positive and negative values of $S_z$, respectively. **(C)** Color-mapped distribution of the $z$ component of Berry curvature $\Omega_z$ on the Fermi surface from both chiral structures. The direction of $D_{zz}$ ($\propto \frac{\partial \Omega_z}{\partial k_z}$) is indicated by green arrows. **(D)** Spin texture $S_z$ of both chiral structures calculated in the $k_y = 0$ plane (i.e., the $k_x$–$k_z$ plane) in the full BZ.

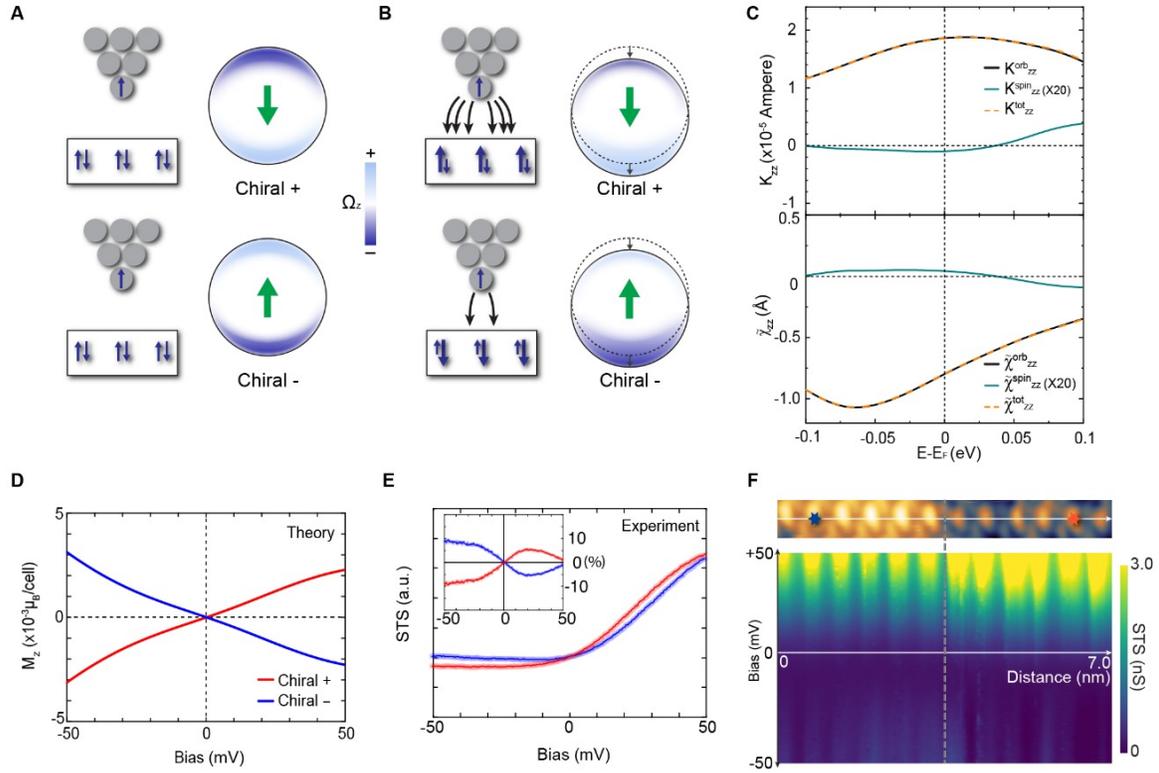

**Figure 4. Berry Curvature Dipole and Induced Magnetization. (A)** Sketch of BCD in reciprocal space (green arrows) and magnetization in real space (blue arrows) for + and – chiral structure without electric current. The BCD pattern does not lead to any net magnetization in the absence of an electric current. **(B)** Change of BCD and magnetizations with electric current. The shift of the Fermi surface results in an imbalance of Berry curvature that is associated with a real-space magnetization that can be detected by a magnetized probe. **(C)** (upper) Calculated orbital, spin, and total gyrotropic tensor element $K_{zz}$ as a function of energy near the Fermi level. (lower) Calculated orbital, spin, and total current-induced magnetization $\tilde{\chi}_{zz}$ as a function of energy near the Fermi level. Both $K_{zz}$ and $\tilde{\chi}_{zz}$, which are related by Eq. (2), are calculated for the – chiral structure, and the spin components are exaggerated 20 times. **(D)** Estimated magnetization $M_z$ for two chiral structures using the calculated total $\tilde{\chi}_{zz}$ and the experimental current density $J_z$. **(E)** Spin-polarized tunneling spectra from two chiral domains (average of +1 and –1 mV normalization bias to minimize the uncertainty in tip height). The red and blue curves are obtained above the Co atoms shown as red and blue stars in (F) respectively. (inset) Same spectra after removing the chirality-averaged signal, shown as a percentage. **(F)** (upper) SP-STM topography where the spectra are taken. (lower) Color-mapped spin-polarized tunneling spectra across a structural chiral domain boundary (vertical dashed line).

20